\documentclass[twocolumn,amsmath,amssymb,prl,superscriptaddress,longbibliography]{revtex4-1}

\def\vec{\mathbf}

\usepackage{titlesec}
\usepackage{xcolor}
\usepackage[citecolor=blue,colorlinks=true,linkcolor=blue]{hyperref}
\usepackage{epsfig}
\usepackage{color}
\usepackage{graphicx} 
\usepackage{dcolumn}  

\newcommand{\be}{\begin{equation}}
\newcommand{\ee}{\end{equation}}

\def\mc{\mathcal}
\def\bs{\boldsymbol}

\begin{document}

\title{Heavy-mass magnetic modes in pyrochlore iridates\\
       due to dominant Dzyaloshinskii-Moriya interaction}
%

\author{Ravi Yadav}
\affiliation
{Institute for Theoretical Solid State Physics, IFW Dresden, Helmholtzstr. 20, 01069 Dresden, Germany}

\author{Manuel Pereiro}
\affiliation
{Department of Physics and Astronomy, Uppsala University, Uppsala 751 20, Sweden}

\author{Nikolay A.~Bogdanov}
\affiliation
{Institute for Theoretical Solid State Physics, IFW Dresden, Helmholtzstr. 20, 01069 Dresden, Germany}

\author{Satoshi Nishimoto}
\affiliation
{Institute for Theoretical Solid State Physics, IFW Dresden, Helmholtzstr. 20, 01069 Dresden, Germany}

\author{Anders Bergman}
\affiliation
{Department of Physics and Astronomy, Uppsala University, Uppsala 751 20, Sweden}
 
\author{Olle Eriksson}
\affiliation
{Department of Physics and Astronomy, Uppsala University, Uppsala 751 20, Sweden}

\author{Jeroen van den Brink}
\affiliation
{Institute for Theoretical Solid State Physics, IFW Dresden, Helmholtzstr. 20, 01069 Dresden, Germany}
\affiliation{Department of Physics, Technical University Dresden, 01062 Dresden, Germany}

\author{Liviu Hozoi}
\affiliation
{Institute for Theoretical Solid State Physics, IFW Dresden, Helmholtzstr. 20, 01069 Dresden, Germany}

\begin{abstract}
Materials with strong spin-orbit interactions are presently a main target in the search for
systems with novel magnetic properties.
Magnetic anisotropies can be very large in such compounds,
ranging from strongly frustrated Kitaev exchange in honeycomb iridates and the associated
spin-liquid states to robust antisymmetric couplings in square-lattice Sr$_2$IrO$_4$.
Here we predict from {\it ab initio} quantum chemistry calculations that another highly unusual
regime is realized in pyrochlore iridium oxides:
the nearest-neighbor Heisenberg interaction can vanish whilst the antisymmetric Dzyaloshinskii-Moriya
exchange reaches values as large as 5 meV,
a result which challenges common notions and existing phenomenological models of magnetic
superexchange.
The resulting spin excitation spectra reveal a very flat magnon dispersion in the Nd- and Tb-based 
pyrochlore iridates, suggesting the possibility of using these modes to store magnetic information.
Indeed the magnetization dynamics indicates that these modes are unable to propagate out of the
excitation region.
\end{abstract}

\date\today
\maketitle

{\it Introduction.\,}
The anisotropy of intersite spin interactions refers to the presence of different coupling
strengths for different spin components $S_i^{\alpha}$, $S_j^{\beta}$ at sites $i$ and $j$
(where $\alpha,\,\beta\!\in\!\{x,y,z\}$) but also to the coupling defined by the cross
product ${\bf S}_i\times {\bf S}_j$.
While the importance of anisotropic exchange has long been recognized in quantum magnetism,
such terms were most often regarded as only small perturbations to the dominant, isotropic
Heisenberg contribution.
However, this view is now changing with the occurrence of both experimental and theoretical 
evidence for large or even prevailing anisotropic interactions in 5$d$ transition-metal
compounds.
In honeycomb Na$_2$IrO$_3$, for example, the intersite spin-coupling anisotropy shows up in the
form of bond-dependent Kitaev interaction terms \cite{kitaev06} --- this symmetric anisotropic 
exchange defines in Na$_2$IrO$_3$ the leading contribution to the effective spin Hamiltonian
\cite{jackeli09,Yamaji_213,katukuri14}.
For nearest-neighbor (NN) IrO$_6$ octahedra in Sr$_2$IrO$_4$, on the other hand, the key
anisotropy is the antisymmetric coupling, also referred to as Dzyaloshinskii-Moriya (DM)
component.
It reaches impressively large values of $\sim$15 meV \cite{jackeli09,Kim12,Perkins14,Bogdanov15},
orders of magnitude larger than in, e.\,g., the isostructural `214' cuprates 
\cite{DM_zhang_rice_1991,aharony_1992,yildirim_1995}.

The bending of the Ir-O-Ir links is sizable in Sr$_2$IrO$_4$, with bond angles of 157$^{\circ}$.
The departure from straight Ir-O-Ir bonds is, however, even more pronounced in the so called
`227' pyrochlore iridates $R_2$Ir$_2$O$_7$, where $R$ stands for a trivalent rare-earth ion:
Ir-O-Ir bond angles as low as 119$^{\circ}$ have been determined in this family of iridium
oxide compounds \cite{disseler13}.
For such a geometry, the isotropic exchange is presumably strongly reduced as compared to the
case of straight bonds in Ba$_2$IrO$_4$ \cite{Katukuri14b,Perkins14} or of moderate flexure as 
in Sr$_2$IrO$_4$ \cite{jackeli09,Kim12,Perkins14,Bogdanov15}.
But the antisymmetric DM coupling is likely to pick up values comparable to those in
Sr$_2$IrO$_4$ --- 
the very atypical situation may arise where the DM interaction $D$ even exceeds the
Heisenberg $J$.
The obvious question is then what kind of unconventional magnetic orders and excitations can
emerge in this case.
With $D/J$ ratios of $\sim$1 and ferromagnetic (FM) $J$'s, unusual skyrmionic excitations
  were recently predicted  for the kagom\'{e}-like [111] planes of the pyrochlore lattice
\cite{Pereiro2014}.
For particular regions in the parameter space those skyrmionic states are stable even at room
temperature, which has potential for applications to logic devices or for data storage.

Few-layer [111] slab structures of 227 iridates are predicted to display in the 
magnetically ordered all-in/all-out (AIAO) configuration anomalous Hall effect \cite{Yang14},
Chern insulator states \cite{Hu15}, emergent domain-wall metallic states \cite{Youhei},
and topological magnon bands \cite{Laurell16}.
Experimental indications of AIAO magnetic order have been recently reported for Nd$_2$Ir$_2$O$_7$
\cite{Tomiyasu12,Ueda15,Tian15}, Tb$_2$Ir$_2$O$_7$ \cite{Lefrancois15}, Eu$_2$Ir$_2$O$_7$
\cite{Sagayama13}, and Sm$_2$Ir$_2$O$_7$ \cite{227_donnerer_2016}.
However, little is known with respect to the strength of the intersite magnetic couplings in
227 iridates.
To access this kind of information, essential for a detailed understanding of the 227 iridate
pyrochlores, we here employ many-body quantum chemistry {\it ab initio} methods.
Subsequently we use classical Monte Carlo simulations to determine the magnetic ground state 
of the resulting effective spin Hamiltonian and perform calculations within linear spin-wave
theory on noncollinear configurations to derive the
magnon excitation spectra.
We find that variation of the Ir-O-Ir bond angles, from $>$130$^{\circ}$ in Sm$_2$Ir$_2$O$_7$
and Eu$_2$Ir$_2$O$_7$ \cite{Taira2001} to 119$^{\circ}$ in Nd$_2$Ir$_2$O$_7$ \cite{disseler13},
induces dramatic modifications of the $D/J$ ratio, from $D/J\!<\!0.2$ in Sm$_2$Ir$_2$O$_7$ and 
Eu$_2$Ir$_2$O$_7$ to $D/|J|\!\gtrsim\!3$ for bond angles around 120$^{\circ}$.
The latter result, $D/J\!>\!3$, is outstanding and shows that, e.g., the topological magnon bands
recently predicted for $D/J\!>\!0.7$ \cite{Laurell16} can be indeed realized in pyrochlore-derived
bilayer nanostructures.
These very large $D/J$ values and the possibility of tuning the sign of the Heisenberg coupling by
changing the bond angles single out the 227 iridates, either as bulk or thin films, as an ideal
playground for the study of DM-driven quantum magnetism, a research area that holds promise both
from the perspective of fundamental science \cite{Yang14,Hu15,Youhei,Laurell16,Pereiro2014}
and at the level of device applications \cite{Pereiro2014}.

\begin{figure}[b]
\includegraphics[width=8.5cm]{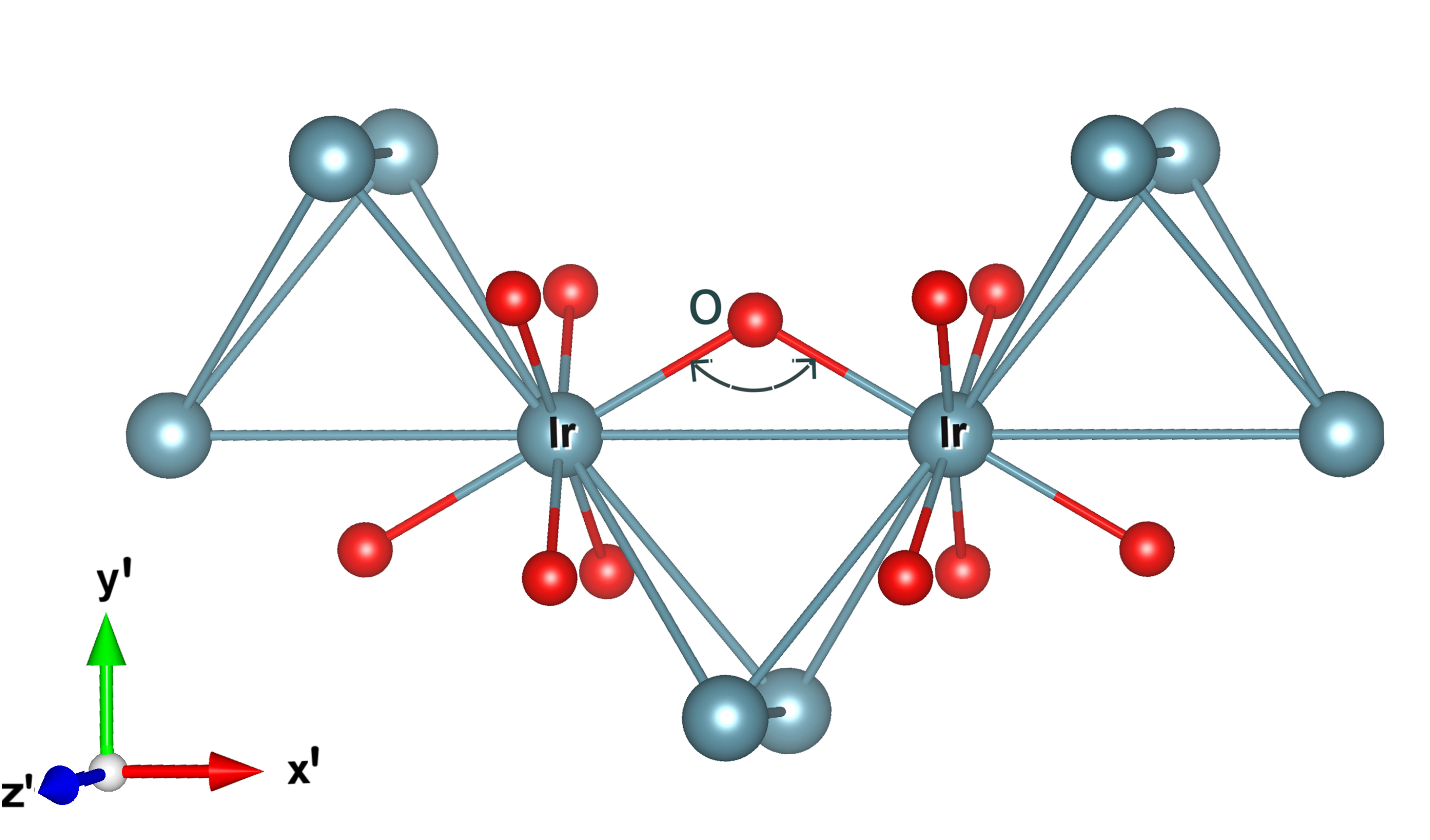}
\caption{
Ir- and O-ion distribution within the $R_2$Ir$_2$O$_7$ pyrochlore lattice.
The network of corner-sharing Ir$_4$ tetrahedra (larger blue spheres) and the O ligands around 
two NN Ir sites (smaller red spheres) are emphasized.
Two adjacent IrO$_6$ octahedra share a single O ion.
The choice of the local coordinate system $\{x',y',z'\}$ is discussed in the text.
}
\label{fig1}
\end{figure}

{\it Electronic-structure calculations.\,}
The Ir ions form a network of corner-sharing regular tetrahedra in the pyrochlore iridates, as
sketched in Fig.~\ref{fig1}.
The lattice constant ($a$) of the {\sl fcc} unit cell varies in the range of 1--1.05 nm in these
compounds.
In addition to the $a$ parameter, one other important structural detail is the fractional
coordinate $x$ of the O anion at the 48$f$ site \cite{Taira2001}, which determines the amount
of trigonal distortion of the O$_6$ octahedral cage around each Ir cation:
cubic, undistorted O octahedra are realized only for $x\!=\!x_c\!=\!5/16$.
%
The $x$ parameter is always larger than $x_c$ in the 227 iridates, which translates into trigonal
squashing of the ligand cages and Ir-O-Ir bond angles $\theta\!<\!\theta_c\!=\!141.1^{\circ}$.
The strong bending of Ir-O-Ir links and the angle $\theta$ are highlighted in Fig.~\ref{fig1}.

The Ir 5$d$-shell formal oxidation state is 4$+$, with one hole in the $t_{2g}$ sublevels.
In the presence of strong spin-orbit coupling (SOC), the latter are split into fully occupied
$j_{\mathrm{eff}}\!\approx\!3/2$ and half-filled $j_{\mathrm{eff}}\!\approx\!1/2$ spin-orbit
states \cite{jackeli09}.
To determine the strength of the various exchange interactions between such NN 1/2-pseudospins, we
carried out many-body quantum chemistry calculations for fragments of two corner-sharing IrO$_6$
octahedra.
We used energy-consistent relativistic core-level pseudopotentials plus valence basis functions
of quadruple-zeta quality for Ir, all-electron quintuple-zeta basis sets for the bridging O
ligand, and triple-zeta basis sets for the other ten O ions of the two-octahedra block (see
Appendix for further details).
Multiconfigurational wave functions were first obtained by complete-active-space 
self-consistent-field (CASSCF) calculations \cite{Helgaker2000} for this [Ir$_2$O$_{11}$]
fragment.
The variational optimization was performed for an average of the lowest nine singlet and nine
triplet states, mostly of $t_{2g}^5$--$t_{2g}^5$ character. 
However, since intersite excitations of $t_{2g}^4$--$t_{2g}^6$ type contribute with finite 
weight to the CASSCF wave functions, the CASSCF calculations account not only for NN
$t_{2g}^5$--$t_{2g}^5$ direct exchange but also for $t_{2g}^4$--$t_{2g}^6$ superexchange
processes.
Further, additional superexchange paths involving the Ir 5$d$ $e_g$ or/and the oxygen $2p$
orbitals show up in the subsequent multireference configuration-interaction (MRCI) computations
\cite{Helgaker2000}.
The latter include all possible single and double electron excitations out of the Ir 5$d$
$t_{2g}$ and bridging-ligand $2p$ shells.

All nine $t_{2g}^5$--$t_{2g}^5$ triplet and low-lying nine singlet states enter the spin-orbit
treatment, both at the CASSCF and MRCI level.
The mapping of the {\it ab initio} quantum chemistry data onto the effective spin model implies
however only the lowest four spin-orbit states, associated with the different possible couplings
of two 1/2 pseudospins.
The other 32 spin-orbit states in this manifold involve $j_{\mathrm{eff}}\!\approx\!3/2$ to 
$j_{\mathrm{eff}}\!\approx\!1/2$ charge excitation and lie at $\gtrsim$0.5 eV higher energy
\cite{jackeli09,Hozoi14}.

\begin{table}[!t]
\caption{
Intersite effective spin couplings (meV) as derived from spin-orbit MRCI calculations in various
227 iridates.
The form of the effective spin Hamiltonian and the choice of the coordinate system are described
in the text.
}
\begin{tabular}{|l|c|r|r|r|r|}
\hline
\hline
                       &       &      &     &              &             \\[-0.25cm]
                       &$\measuredangle$(Ir-O-Ir)
                               &$J$   &$D$  &$\Gamma_{x'x'}$ &$\Gamma_{y'y'}$\\
\hline
                       &       &      &     &              &   \\[-0.20cm]
Sm$_2$Ir$_2$O$_7$      &132    &22.4  &2.2  &--1.7          &--0.58\\[0.04cm]
Eu$_2$Ir$_2$O$_7$      &131    &19.3  &2.4  &--2.0          &--0.49\\[0.04cm]
Y$_2$Ir$_2$O$_7$       &130    &18.8  &2.6  &--2.2          &--0.46\\[0.04cm]
Er$_2$Ir$_2$O$_7$      &129    &13.5   &3.1  &--2.4         &--0.35\\[0.04cm]
Lu$_2$Ir$_2$O$_7$      &126    &8.9    &3.6  &--2.9         &0.31\\[0.04cm]
Tb$_2$Ir$_2$O$_7$      &121    &--1.5 &4.7  &--3.4         &--0.06\\[0.04cm]
Nd$_2$Ir$_2$O$_7$      &119    &--1.6 &5.1  &--3.8         &0.02\\[0.04cm]
\hline
\hline
\end{tabular}
\label{QC_spin_couplings}
\end{table}

{\it Magnetic model Hamiltonian.\,}
A pair $\langle ij\rangle$ of NN $j_{\mathrm{eff}}\!\approx\!1/2$ sites is magnetically described
by the following bilinear effective spin Hamiltonian\,:
\be\label{SpinHam_2sites}
\mc{H}^{i,j}_{\mathrm{eff}} =
   J\, \tilde{\vec{S}}_i \cdot \tilde{\vec{S}}_j
   +{\vec D} \cdot {\tilde {\bf S}}_i\times {\tilde {\bf S}}_j
   +\tilde{\vec{S}}_i \cdot \bar{\bar{\bs{\Gamma}}} \cdot \tilde{\vec{S}}_j\,
   +\mu_{\rm B}\!\sum_{k\!=\!i,j} \vec{h} \cdot \bar{\bar{\vec{g}}}_k \cdot \tilde{\vec S}_k\,,
\ee
where $\tilde{\vec{S}}_i$, $\tilde{\vec{S}}_j$ are pseudospin-1/2 operators, $J$ is the 
isotropic Heisenberg coupling, $\vec{D}$ stands for the antisymmetric DM exchange,
$\bar{\bar{\bs{\Gamma}}}$ is a symmetric traceless second-rank tensor defining the symmetric
anisotropic interaction, and the last term describes the coupling to an external magnetic field
$\vec{h}$.
The anisotropy of the latter component is characterized by the $\bar{\bar{\vec g}}$ tensor.

Since a block of two NN IrO$_6$ octahedra has $C_{2v}$ point-group symmetry in 227 pyrochlore
iridates, with two-fold rotational symmetry around the Ir-Ir axis (see Fig.\;\ref{fig1}), a  
convenient reference system is for many purposes a local frame having one of the coordinates
along the line defined by those two Ir sites.
Such a coordinate system, $\{x',y',z'\}$, is used in Fig.\;\ref{fig1} and form part of the
following discussion, with the $x'$ axis taken along the Ir-Ir link and $z'$ perpendicular to
the triangular plaquette formed by the two Ir NN's and the bridging ligand.
In this frame, all off-diagonal elements $\Gamma_{\!\alpha\beta}$ are 0 and we can then write
\cite{katukuri14,Bogdanov15}
\be\label{Gamma}
\bar{\bar{\bs{\Gamma}}} =
\left(
\begin{array}{ccc}
\Gamma_{x'x'}    &0           &0 \\
0              &\Gamma_{y'y'} &0 \\
0              &0           &-\Gamma_{x'x'}-\Gamma_{y'y'}
\end{array}
\right)\,.
\ee
The $C_{2v}$ symmetry further simplifies the form of the DM vector, with $\vec{D}\!=\!(0,0,D)$.
It is the effective spin model described by Eqs.~(\ref{SpinHam_2sites}) and (\ref{Gamma}) onto
which we map the {\it ab initio} quantum chemistry data for the spin-orbit, two-octahedra
$d^5$--$d^5$ states (see \cite{Bogdanov15,Yadav16} and Appendix for details).
Intersite magnetic couplings based on spin-orbit MRCI calculations and such a mapping scheme
are provided in Table~\ref{QC_spin_couplings}.

In contrast to usual plots in which various quantities are pictured as function of the $R$-ion
radius \cite{Matsuhira2011}, we focus in Table~\ref{QC_spin_couplings} on overall trends when
gradually reducing the Ir-O-Ir bond angle. 
In addition to the lanthanide-based 227 iridates, we also include in the table the Y-iridate,
Y$_2$Ir$_2$O$_7$.

\begin{figure}[b]
\includegraphics[width=8.3 cm]{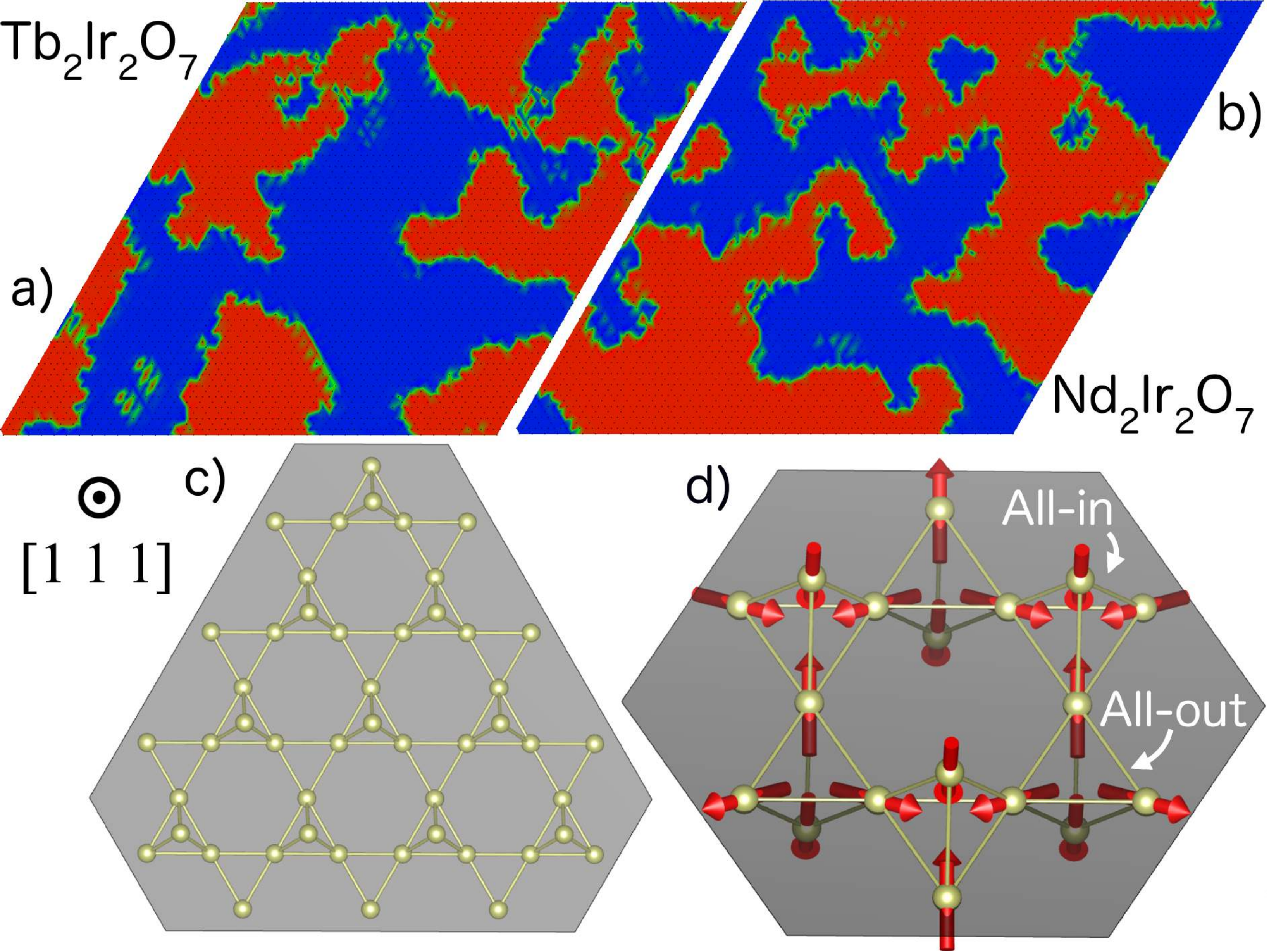}
\caption{
Ground-state spin structures in 227 iridates.
Computational results for Tb227 and Nd227 are provided in (a) and (b), respectively, for planar
atomic configurations of 40$\times$40 unit cells normal to the [1\,1\,1] crystallographic direction.
The blue and red sectors represent different orientations of the $z$ components of the Ir magnetic moments,
parallel or antiparallel to the $z$ axis.
The transition region between those sectors is highlighted in green.
A kagom\'{e} Ir plane onto which the magnetic momens in (a) and (b) are mapped is sketched in (c) while
(d) displays the AIOA spin arrangement found for the other iridates studied here.
}
\label{fig2}
\end{figure}

From the MRCI results listed in Table~\ref{QC_spin_couplings} it is seen that the $\Gamma$ coupling
parameters are never the largest interactions.
The overall tendencies for the other exchange couplings are obvious: from $J$ and $D$ values in
the range $J\!>\!20$ and $D\!>\!2$ meV for the largest Ir-O-Ir bond angles,
one arrives at $J\!\approx\!-1.5$ and $D\!\approx\!5$ meV for bond angles of 119--121$^{\circ}$ in
Nd$_2$Ir$_2$O$_7$ and Tb$_2$Ir$_2$O$_7$.
In other words, the $D/J$ ratio changes by a factor of $\approx$30 along the series, with the
highly unusual situation of having a DM interaction $D$ a few times larger than the Heisenberg
$J$ in Nd$_2$Ir$_2$O$_7$ and Tb$_2$Ir$_2$O$_7$.
What is more interesting, $J$ changes sign for Ir-O-Ir angles of 121--125$^{\circ}$.
This suggests that even larger $D/J$ ratios are in principle attainable by
using strain in thin films of 227 iridates or pressure, to reach angles in the 121--125$^{\circ}$
interval for which $J\!\rightarrow\!0$.
%

\begin{table}[!t]
\caption{
Intersite effective spin couplings (meV) at different levels of theory for the smallest
(Nd$_2$Ir$_2$O$_7$) and largest (Sm$_2$Ir$_2$O$_7$) Ir-O-Ir bond angles reported in 227
iridates \cite{Taira2001,disseler13}.
}
\begin{tabular}{|c|l|  r  r  r  r |}
\hline
\hline\\[-0.30cm]
227 system     &Method &{\hskip 0.25cm} $J$ &{\hskip 0.25cm} $D$ &{\hskip 0.25cm} $\Gamma_{x'x'}$  &{\hskip 0.25cm} $\Gamma_{y'y'}$\\
\hline\\[-0.20cm]
Sm$_2$Ir$_2$O$_7$
              & MRCI      &$22.4$   &$2.2$  &$-1.7$         &$-0.6$ \\[0.08cm]
              & CASSCF    &$11.5$   &$1.2$  &$-0.9$         &$0.2$ \\  [0.08cm]
              & ROHF      &$-1.2$   &$0.3$  &$-0.2$         &$1.2$ \\   [0.08cm]
\hline\\[-0.20cm]
Nd$_2$Ir$_2$O$_7$
              & MRCI      &$-1.6$   &$5.1$  &$-3.8$         &$0.02$ \\ [0.08cm]
              & CASSCF    &$-1.8$   &$4.6$  &$-2.3$         &$-0.3$ \\ [0.08cm]
              & ROHF      &$-0.4$   &$0.4$  &$-0.2$         &$0.2$ \\  [0.08cm]
\hline
\hline
\end{tabular}
\label{QC_diff_approx}
\end{table}

{\it Emergent domain-wall state.\,}
Starting from the effective coupling parameters displayed in Table~\ref{QC_spin_couplings},
the magnetic ground-state configurations of the series of 227 iridates were computed in a
two-stage approach.
In a first step, classical Monte Carlo simulated annealing~\cite{simanneal} was used to
`thermalise' the system, at a temperature of 0.001 K.
Data sampling was subsequently performed by using an atomistic spin dynamics (ASD) method, as
implemented in the UppASD code (see \cite{asd} and Appendix for further information).
With the exception of Tb227 and Nd227, the atomic spin moments of the resulting ground-state
magnetic structures follow a AIAO pattern, see Fig.~\ref{fig2}.(d).
Plots for the orientation of the $z$ components of the magnetic moments, ``$+z$'' and ``$-z$'',
at sites within a plane perpendicular to the [1\,1\,1] crystallographic direction \cite{Taira2001}
are provided in Figs.~\ref{fig2}.(a)-(b) for both Tb$_2$Ir$_2$O$_7$ and Nd$_2$Ir$_2$O$_7$.
The spin configuration is collinear and FM in each of the blue ($+z$) and red ($-z$) sectors and
there is a transition region (shown in green color) where the moments rotate continuously from
parallel orientation in blue to antiparallel in red, giving rise to an {\it emergent
domain-wall state}~\cite{Youhei}.
The arrangement of Ir sites within a  kagom\'{e} plane of the pyrochlore lattice is depicted in
Fig.~\ref{fig2}.(c).
Such a plane has been chosen for the plots in Figs.~\ref{fig2}.(a)-(b).
The degenerate magnetic ground state associated with the domain-wall stucture can be destroyed
by small magnetic fields applied along the [1\,1\,1] crystallographic direction.

We note that the $R$-ion magnetic sublattice, believed to actively contribute to the realization
of AIAO order in $R_2$Ir$_2$O$_7$ pyrochlores \cite{Tomiyasu12,Ueda15,Tian15,Lefrancois15}, is 
not accounted for in our computations.
In other words, for a given $D/J$ ratio, our simulations would somewhat overestimate the tendency
towards the formation of domain-wall states.
But this instability is there, `nearby' and further modification of the $D/J$ ratio through 
appropriate engineering of the Ir-O-Ir angle, via either strain or pressure, should drive some of
these materials towards that kind of regime.
According to our MRCI results, the most `responsive' should be the Tb227, Nd227, and Lu227
systems.

\begin{figure}[t]
\includegraphics[width=8.3 cm]{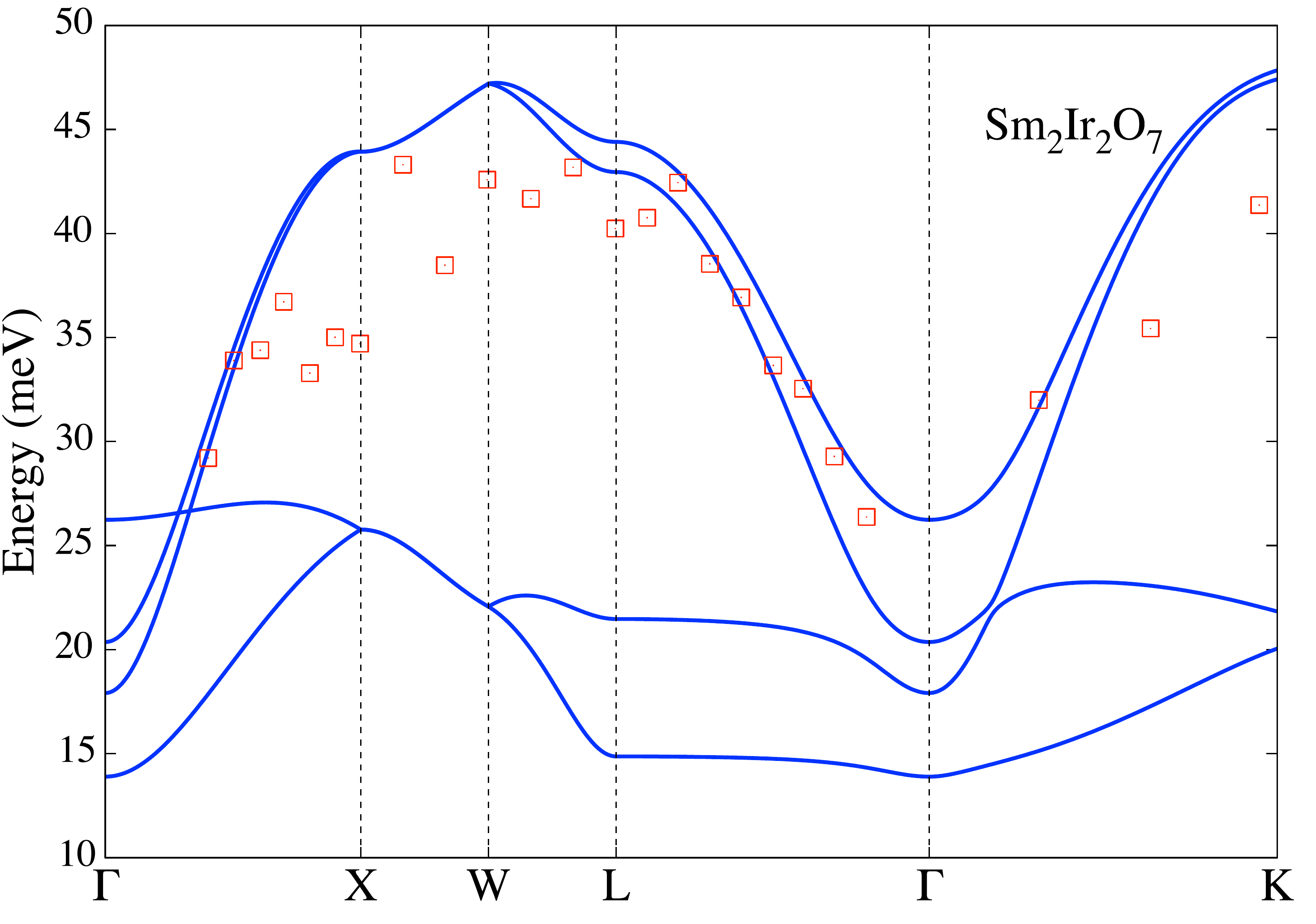}
\caption{
Adiabatic magnon spectrum of Sm$_2$Ir$_2$O$_7$ (blue lines).
The empty squares indicate the experimental data of Ref.~\cite{227_donnerer_2016}.
%
The four different branches are related to the four inequivalent Ir sites in the unit cell.
}
\label{fig3}
\end{figure}

{\it Spin dynamics.\,}
It is instructive to confront these theoretical results and predictions with measurements on
the properties of the magnetic system in pyrochlore iridates.
In particular, spin excitation spectra have been recently measured for Sm$_2$Ir$_2$O$_7$
\cite{227_donnerer_2016} using resonant inelastic x-ray scattering (RIXS).
For comparison, we show in Fig.~\ref{fig3} the adiabatic magnon spectra of Sm$_2$Ir$_2$O$_7$ as
calculated for a noncollinear magnetic ground state, using linear spin wave theory (LSWT) and
the UppASD computer implementation.
The spectra were obtained on the basis of the effective coupling parameters of Table~\ref{QC_spin_couplings}
and a moment 0.75 $\mu_{\mathrm B}$ at each Ir site.
It is seen that the theoretical magnon spectrum agrees well with the experimental data reported
in Ref.~\cite{227_donnerer_2016}.
The gap of the acoustical branch is due to having both large DM and symmetric anisotropic couplings.
Donnerer {\it et al.}~\cite{227_donnerer_2016} found that a rather good fit of the experimental
dispersion is obtained by LSWT and $J\!=\!27$, $D\!=\!4.9$ meV.
Our {\it ab initio} results for Sm$_2$Ir$_2$O$_7$, $J\!\approx\!22.5$ and $D\!=\!2.2$ meV (see
Table~\ref{QC_spin_couplings}), are somewhat on the lower side, as compared to the outcome of the
LSWT fits in Ref.\,\cite{227_donnerer_2016}.
However, analysis of the quantum chemistry data at different levels of approximation --- restricted
open-shell Hartree-Fock (ROHF) \cite{Helgaker2000}, CASSCF, and MRCI (see Table~\ref{QC_diff_approx})
--- shows that while the corrections brought by MRCI to CASSCF are important for large Ir-O-Ir
angles, they are minor for the lowest bond angles, in the range of only tenths of a meV.
This implies that our prediction of small, FM $J$ values for strongly bent Ir-O-Ir links in 
Nd$_2$Ir$_2$O$_7$ and Tb$_2$Ir$_2$O$_7$ is solid.
MRCI results in good agreement with experimental data were earlier reported using the same quantum
chemistry approach for corner-sharing IrO$_6$ octahedra with Ir-O-Ir bond angles of 180$^{\circ}$ 
in Ba$_2$IrO$_4$ \cite{Katukuri14b}, 157$^{\circ}$ in Sr$_2$IrO$_4$ \cite{Bogdanov15}, and
140$^{\circ}$ in CaIrO$_3$ \cite{nikolay_cairo_PRB}.
Remarkably large variations of the anisotropic interactions as function of bond angles were also
computed for edge-sharing octahedra in honeycomb iridates \cite{Nishimoto16}, with the Kitaev
coupling evolving between $\approx$0 meV at 90$^{\circ}$, an {\it ab initio} result which challenges
present superexchange models \cite{jackeli09}, to $\approx$25 meV for 100$^{\circ}$.

\begin{figure}[!t]
\includegraphics[width=8.3 cm]{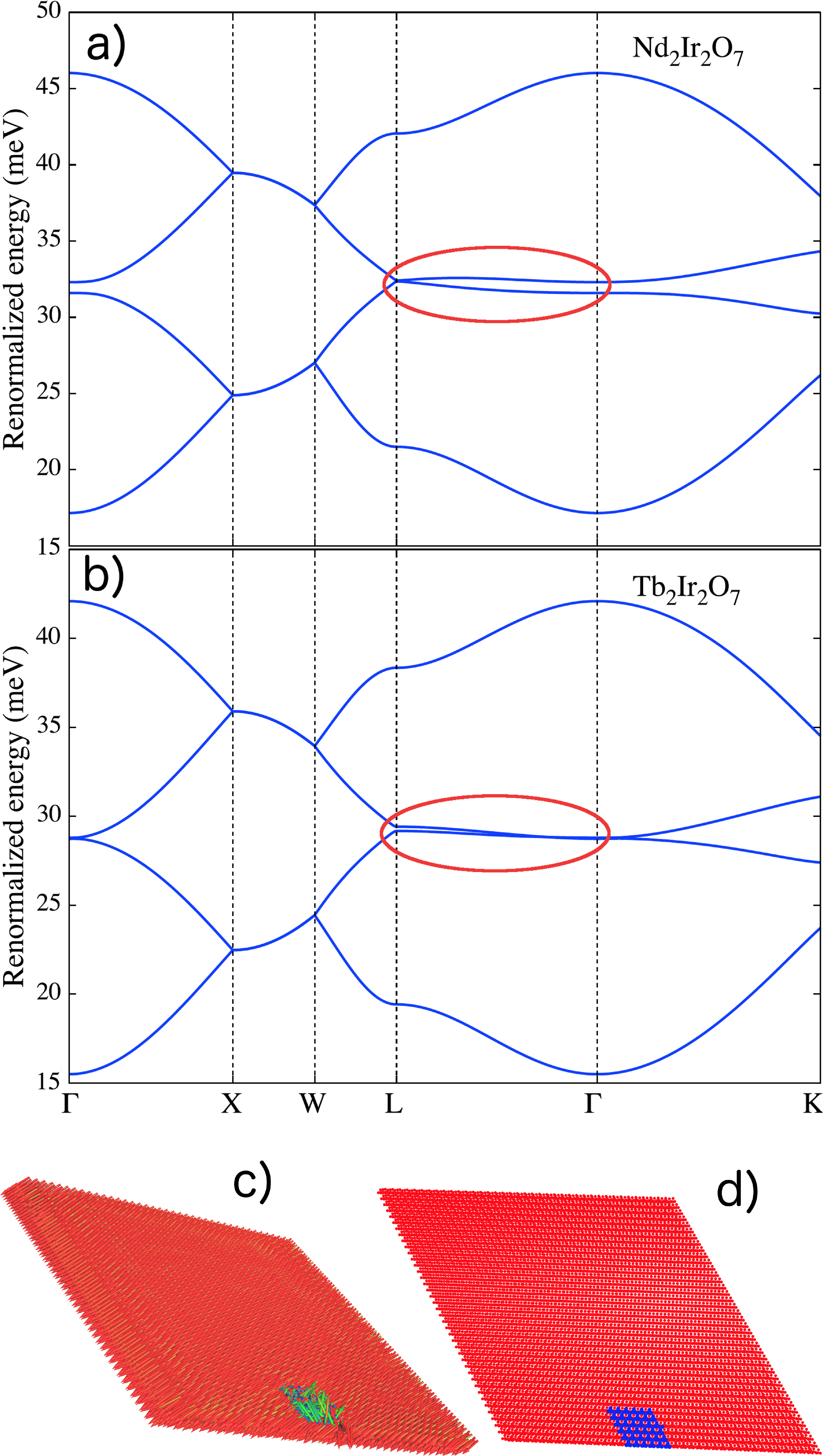}
\caption{
Adiabatic magnon spectra for (a) Nd227 and (b) Tb227.
The region with almost flat bands is indicated with a red ellipse.
(c) Snapshot after applying a time-dependent external magnetic field in Nd227.
The external perturbation is applied locally, within a region of approximately
6$\times$6$\times$1 unit cells; the sampled volume is 40$\times$40$\times$3 unit cells.
`Perturbed' spins are indicated in green color.
(d) Time-averaged magnetic moments.
Magnon activity is indicated in blue color.
}
\label{fig4}
\end{figure}

Adiabatic magnon spectra computed for other pyrochlore iridates of the 227 family are provided
in Fig.~\ref{fig4}.(a)-(b) and in Supplementary Information.
The spectra of Sm227, Eu227, Er227, Lu227, and Y227 display very similar features, see Figs.~\ref{fig3}
and S1, as all these compounds possess the same type of AIAO magnetic order.
Nd227 and Tb227, on the other hand, exhibit an instability towards the formation of emergent
domain-wall states.
Magnon spectra for the case of collinear FM ground states are displayed for these two materials in
Fig.~\ref{fig4}.
Interestingly, both spectra feature very flat (nearly degenerate) bands at intermediate energies
along the L$\rightarrow\!\Gamma$ direction [red ellipses in Figs.~\ref{fig4}.(a)-(b)].
A nearly vanishing magnon dispersion in the L$\rightarrow\!\Gamma$ sector brings the prospect
of using magnetic excitations which reverse the orientation of the magnetic moments for storing
permanent magnetic information.
Along this line of thinking, we show in Fig.~\ref{fig4}.(d) the Ir-site magnetic moments averaged
over a simulation time of 82 ps.
The region in red color indicates an area where the Ir moments are constant in time, pointing out
of the kagom\'{e} [111] plane, while blue color illustrates a strong magnetic perturbation that
reverses the magnetization direction.
The evolution in time of the spin system was modeled according to the Landau-Lifshitz-Gilbert (LLG)
equation of motion, after the system was thermalised at a very low temperature ($10^{-3}$ K).
A time-dependent external magnetic field with a frequency corresponding to the energy range in
which the Nd227 magnon spectra are flat (32.34 meV) was applied in the simulation, within a limited
domain of 6$\times$6$\times$1 unit cells. 
Moreover, the external field was applied along the [1\,1\,1] direction, which corresponds to the
direction L$\rightarrow\!\Gamma$ in reciprocal space.
As shown in Fig.~\ref{fig4}.(d), the magnetic perturbation is not able to propagate out of the
excitation region.
In other words, any perturbation that reverses the orientation of the moments will stay localised
in this area, without propagation or dissipation via spin-waves.

{\it Conclusions.\,}
The microscopic mechanism through which spin-orbit interactions give rise to anisotropic exchange
in magnetic insulators represents a longstanding problem.
With regard to the antisymmetric anisotropy, also referred to as the Dzyaloshinskii-Moriya term, 
a large body of theoretical studies focused so far on extensions of the Hubbard model and layered
square-lattice materials such as La$_2$CuO$_4$ \cite{DM_zhang_rice_1991,aharony_1992,yildirim_1995}
and Sr$_2$IrO$_4$ \cite{jackeli09,Kim12,Perkins14}.
As a rule of thumb, $D\!\ll\!J$, with $D$ values three orders of magnitude smaller than $J$ in
cuprates \cite{yildirim_1995}.
More recently, Dzyaloshinskii-Moriya interaction strengths of 20--25\% of the Heisenberg $J$ have
been estimated in Sr$_2$IrO$_4$ \cite{jackeli09,Kim12,Perkins14,Bogdanov15}.
Here we employ {\it ab initio} quantum-chemistry many-body techniques to reveal a singular situation
in which the Dzyaloshinskii-Moriya coupling parameter $D$ outstrips the isotropic Heisenberg 
interaction.
This is achieved according to our calculations in the pyrochlore iridates Nd$_2$Ir$_2$O$_7$
and Tb$_2$Ir$_2$O$_7$, displaying strongly bent Ir-O-Ir links with bond angles of less than
125$^{\circ}$.
%
  We further show that such a regime is prone to the formation of emergent domain-wall states.
In our computations
both Nd$_2$Ir$_2$O$_7$ and Tb$_2$Ir$_2$O$_7$ feature magnon dispersions that are almost zero
along specific directions. 
This points to the potential use of magnetic excitations which reverse the orientation of the
magnetic moments to store magnetic information.
Indeed, the calculated magnetization dynamics shows that these excited magnetic modes are unable
to propagate out of the excitation region.

 \

 \

{\it Acknowledgements.\,} 
Calculations were performed at the High Performance Computing Center (ZIH) of the Technical
University Dresden (TUD) and at the computational facilities of C3SE at Chalmers University
of Technology in Gothenburg, under a project administrated by SNIC/SNAC.
R.\,Y., S.\,N., and L.\,H.~thank U.~Nitzsche for technical assistance.
L.\,H.~thanks P.\,Fulde and U.\,R{\"o}{\ss}ler for discussions.
We acknowledge financial support from the German Science Foundation (Deutsche Forschungsgemeinschaft,
DFG --- SFB-1143 and HO-4427/2), the KAW Foundation (projects 2013.0020, 2012.0031),
the Swedish Research Council (VR), and eSSENCE.

 \

 \
 
 \ 

\appendix
\noindent
{\bf Appendix}

\noindent
{\it Ab initio calculations.\,} As concerns the quantum chemistry calculations, the material model consists of two IrO$_6$ octahedra sharing one ligand, with bond lengths and
bond angles as determined experimentally \cite{Taira2001,disseler13,Lefrancois15}.
A similar material model has been used in earlier work to establish the strengths of the Kitaev and
Heisenberg effective couplings and their dependence on bond angles for edge-sharing octahedra in
honeycomb-lattice iridates \cite{katukuri14}. We applied energy-consistent relativistic pseudopotentials along with basis sets of quadruple-zeta 
quality\,\cite{Figgen09} for the two Ir sites of the reference octahedra. All-electron basis sets of quintuple-zeta quality
were used for the bridging oxygen, while triple-zeta basis functions were used for the remaining oxygen atoms of the reference octahedra\,\cite{Dunning89}.
We further employed two Ir {\it f}\,\,\cite{Figgen09} and four O {\it d}\,\,\cite{Dunning89} polarization functions for the two central Ir ions and the bridging O$_b$ ligand, respectively.

In the intial step, multiconfiguration wave functions were generated by CASSCF calculations where 10 electons (5 at each Ir$^{4+}$ site) and 6 $t_{2g}$ orbitals (3 at each Ir$^{4+}$ site) were considered in 
the active space. The CASSCF optimization was carried out for the low-lying nine singlets and the nine triplets belonging to this manifold. All these states entered the subsequent spin-orbit calculations\,\cite{SOC_molpro}, 
both at the CASSCF and MRCI level. 
In the MRCI treatment, we accounted for single and double excitations from the Ir $t_{2g}$ and bridiging O$_b$ 2{\it p} orbitals\,\cite{Werner88,Knowles92}.
 A Pipek-Mezey localization procedure \cite{Pipek89} was employed for separating the metal 5$d$ and O 2$p$ valence orbitals into different 
groups (e.g., bridging and non-bridging {\it p} orbitals). All calculations were performed with the {\sc molpro} quantum chemistry package\,\cite{Molpro12}. 
A similar computational strategy has been successfully adopted in earlier quantum chemistry studies
\cite{katukuri14,Bogdanov15,Yadav16}.


{\it Mapping procedure.\,}
The lowest four spin-orbit states obtained from the {\it ab initio} quantum chemistry calculations are mapped onto the respective eigenvectors of the effective spin model defined by
Eq.~(\ref{SpinHam_2sites}). 
%
To illustrate the way the mapping procedure is carried out, relevant matrix elements are provided in Tables \ref{abinitioHam} and \ref{modHam} for Sm$_2$Ir$_2$O$_7$. %
Each of the matrix elements $H^{kl}_\mathit{ab\,initio}$ computed at the quantum chemistry
level, see Table \ref{abinitioHam}, is assimilated to the corresponding matrix element
$H^{kl}_{\mathrm{eff}}$ of the effective spin Hamiltonian, see Table \ref{modHam}.
This one-to-one correspondence between {\it ab initio} and effective-model matrix elements
enables the evaluation of all coupling constants in Eq.~(\ref{SpinHam_2sites}).

For $C_{2v}$ symmetry of the two-octahedra [Ir$_2$O$_{11}$] unit, it is convenient to choose
a reference frame with one of the axes along the Ir-Ir link.
The data shown in Tables \ref{abinitioHam} and \ref{modHam} are expressed by using such a
coordinate system, with the $x'$ axis along the Ir-Ir segment and $z'$ perpendicular to the
Ir-O$_b$-Ir plane.
In this framework, the $\bar{\bar{\vec{g}}}$ tensor describing the Zeeman interaction term in
Eq.~(\ref{SpinHam_2sites}) takes the following form at each Ir site $k\in\{i,j\}$~\cite{hill72,oshikawa97}\,:
\be\label{gtensor}
\bar{\bar{\vec{g}}}_{k} =
\left(
\begin{array}{ccc}
g_{x'x'}    & g_{x'y'}          &0 \\
g_{x'y'}              &g_{y'y'} &0 \\
0              &0           &g_{z'z'}
\end{array}
\right)\,.
\ee
Since the Zeeman coupling to an external magnetic field can be also expressed as
$\hat{\cal H}_{\mathrm{Z}}^{i,j}=\mu_{\mathrm{B}} ({\bf L} + g_e {\bf S})\cdot\bf{h}$\,,
all required matrix elements can be easily obtained on the basis of the {\sc molpro} output data,
i.e., the expectation values of the total angular-momentum (${\mathbf L}$) and spin (${\mathbf S}$)
operators (see also Refs.~\cite{Bogdanov15,Yadav16}).
$\Delta_{+}$ and $\Delta_{-}$ stand in Table~\ref{modHam} for
\begin{equation}
\label{delta}
\Delta_{\pm}= 2J+ \Gamma_{x'x'} \pm \sqrt{4D^2+(2J+\Gamma_{x'x'})^2} \,.
\end{equation}

\begin{table*}[t]
\caption{
{\bf Matrix elements of the {\em ab initio} model Hamiltonian (meV) for Sm$_2$Ir$_2$O$_7$, as obtained by spin-orbit
MRCI.}
The two-site singlet and (split) triplet states are labeled $|\tilde s \rangle$ and
\{$|t_{x}\rangle$,\,$|t_{y}\rangle$,\,$|\tilde t_{z}\rangle$\}, respectively.
$|\tilde s\rangle$ and $|\tilde t_{z}\rangle$ are admixtures of `pure' $|0,0\rangle$
and $|1,0\rangle$ spin functions.
}
\begin{tabular}{c c c c c}
\hline
\hline\\[-0.30cm]
$H^{kl}_\mathit{ab\,initio}$

               \hspace{0.4cm}  &$|\tilde s\rangle$ \hspace{0.4cm} &$|\tilde t_{z}\rangle$ \hspace{0.4cm}
                                                                  &$|t_{y}\rangle$ \hspace{0.4cm}
                                                                  &$|t_{x}\rangle$ \\[0.20cm]

$\langle \tilde s|$ \hspace{0.7cm}&$0$             \hspace{0.4cm} &$0$\hspace{0.4cm}

                                                      &$-0.507i\,\mu_B h_{x'}$\hspace{0.4cm}

                                                                        &$-0.296i\,\mu_B h_{y'}$ \\[0.20cm]

$\langle \tilde t_{z}|$\hspace{0.4cm}

                 &$0$\hspace{0.4cm}

                                   &$21.414$  \hspace{0.4cm}       &$-1.294i\,\mu_B h_{x'}$\hspace{0.4cm}

                                                                        &$-2.231i\,\mu_B h_{y'}$ \\[0.20cm]

$\langle t_{y}|$\hspace{0.4cm}

                 &$0.507i\,\mu_B h_{x'}$\hspace{0.4cm}

                                   &$1.294i\,\mu_B h_{x'}$\hspace{0.4cm}

                                                      &$22.773$\hspace{0.4cm}        &$1.896i\,\mu_B h_{z'}$ \\[0.20cm]

$\langle t_{x}|$\hspace{0.4cm}

                 &$0.296i\,\mu_B h_{y'}$   \hspace{0.4cm}           &$2.231i\,\mu_B h_{y'}$\hspace{0.4cm}

                                                      &$-1.896i\,\mu_B h_{z'}$\hspace{0.4cm}

                                                                        &$23.512$ \\
\hline
\hline
\end{tabular}
\label{abinitioHam}
\end{table*}

\begin{table*}[!t]
\caption{
{\bf Matrix form of the effective spin Hamiltonian in the basis of zero-field eigenstates.}
$\Gamma_{2z'\pm x'}$ stands for $2\Gamma_{z'z'} \pm\Gamma_{x'x'}$\,; expressions for the $\Delta_+$ and $\Delta_-$
terms are provided in Eq.~(\ref{delta}).} 
{ \scriptsize
\begin{tabular}{c c c c c}
\hline
\hline\\[-0.30cm]
$H^{kl}_{\mathrm{eff}}$

               \hspace{0.1cm}  &$|\tilde s\rangle$ \hspace{0.1cm} &$|\tilde t_{z}\rangle$ \hspace{0.1cm}
                                                                  &$|t_{y}\rangle$ \hspace{0.1cm}
                                                                  &$|t_{x}\rangle$ \\[0.40cm]

$\langle \tilde s|$ \hspace{0.7cm}&$0$             \hspace{0.1cm} &$0$\hspace{0.1cm}

                                                      &$- \frac{ ih_{x'}(2D g_{x'x'} + g_{x'y'}\Delta_+))}{\sqrt{4D^2 + \Delta_+^2}}$ \hspace{0.1cm}

                                                                      & $- \frac{ ih_{y'}(2D g_{y'y'} - g_{x'y'}\Delta_+))}{\sqrt{4D^2 + \Delta_+^2}}$ \\[0.40cm]

$\langle \tilde t_{z}|$\hspace{0.1cm}

                 &$0$\hspace{0.1cm}

                                   &$\frac{1}{2} \sqrt{4D^2+(2J+\Gamma_{x'x'})^2}$  \hspace{0.1cm}       &$- \frac{ ih_{x'}(2D g_{x'x'} + g_{x'y'}\Delta_-))}{\sqrt{4D^2 + \Delta_-^2}}$ \hspace{0.1cm}

                                                                      & $- \frac{ ih_{y'}(2D g_{y'y'} -g_{x'y'}\Delta_-))}{\sqrt{4D^2 + \Delta_-^2}}$ \\[0.40cm]

$\langle t_{y}|$\hspace{0.1cm}

                 &$ \frac{ ih_{x'}(2D g_{x'x'} + g_{x'y'}\Delta_+))}{\sqrt{4D^2 + \Delta_+^2}}$\hspace{0.1cm}

                                   &$ \frac{ ih_{x'}(2D g_{x'x'} + g_{x'y'}\Delta_-))}{\sqrt{4D^2 + \Delta_-^2}}$ \hspace{0.1cm}

                                                      &$\frac{1}{4} (2J + \Gamma_{2z'+x'} +\sqrt{4D^2+(2J+\Gamma_{x'x'})^2})$  \hspace{0.1cm}        &$ig_{z'z'} h_z'$ \\[0.40cm]

$\langle t_{x}|$\hspace{0.1cm}

                 &$ \frac{ ih_{y'}(2D g_{y'y'} + g_{x'y'}\Delta_+))}{\sqrt{4D^2 + \Delta_+^2}}$  \hspace{0.1cm}          

                                    &$ \frac{ ih_{y'}(2D g_{y'y'} -g_{x'y'}\Delta_-))}{\sqrt{4D^2 + \Delta_-^2}}$

                                                        &$-ig_{z'z'}h_{z'}$\hspace{0.1cm}

                                                                          &$\frac{1}{4} (2J + \Gamma_{2z'-x'} + \sqrt{4D^2+(2J+\Gamma_{x'x'})^2})$ \\
\hline
\hline
\end{tabular}
}
\label{modHam}
\end{table*}

{\it Spin dynamics and LSWT for noncollinear systems.\,}
In order to calculate the ground-state magnetic configurations of the series of 227 iridates,
we used in the initial stage classical Monte Carlo simulated annealing~\cite{simanneal} to thermalise
the system.
Subsequently, data sampling was performed by ASD simulations~\cite{asd}.
During the annealing process the initial temperature was fixed to 500 K and then reduced in the
sequence 500--300--100--50 K to the very last step where the targeted temperature was 0.001 K.
In order to thermalise the system, we used 100,000 Monte Carlo steps for each temperature.

The equation of motion of the classical atomistic spins reads in the LLG frame\,:
\begin{eqnarray}
\frac{\partial s_i}{\partial t}&=&-\frac{\gamma}{1+\alpha_i^2} {\bf s}_i \times [{\bf B}_i+{\bf b}_i(t)]\nonumber\\&-&\frac{\gamma \alpha_i}{s (1+\alpha_i^2)} {\bf s}_i \times \left\{{\bf s}_i \times [{\bf B}_i+{\bf b}_i(t)]\right\} \,,
\end{eqnarray}
where $\gamma$ is the gyromagnetic ratio, $\alpha_i$ denotes a dimensionless site-resolved damping parameter which accounts for the energy dissipation that eventually brings the system into thermal equilibrium, the effective field is calculated as ${\bf B}_i=-\partial \mathcal{H}_\mathrm{eff}/\partial {\bf s}_i$, and temperature ($T$) fluctuations are considered through a random Gaussian-shaped field ${\bf b}_i(t)$. To calculate the ground-state magnetic configurations we assumed a Gilbert damping $\alpha$=0.1, which allows to reach the ground-state in a short period of time. The total simulation time was 320 ps.

The adiabatic magnon spectra were derived within the LSWT frame for a noncollinear ground-state
magnetic structure, following a strategy similar to that described in Ref.~\cite{toth}.
For the case of AIAO order the crystallographic unit cell is commensurate with the magnetic cell
and consequently we defined a local coordinate system that transforms the AIAO configuration into
FM order, by applying a rotation ${\mathcal R}$ on every moment within the crystallographic unit
cell.
This rotation is applied to the spin Hamiltonian described by Eq.~(\ref{SpinHam_2sites}) and
provides\,:
\begin{align}
	\mathcal{H}=&\sum_{ij}\left(\sqrt{\frac{s_i}{2}}(\bar{{\bf u}}_i^\mathrm{T}a_i+{\bf u}_i^\mathrm{T}a^\dagger_i+{\bf v}_i^{\mathrm{T}}(s_i-a^\dagger_i a_i))\right)\mathcal{J}_{ij}\nonumber\\
	&\left(\sqrt{\frac{s_j}{2}}(\bar{{\bf u}}_j^\mathrm{T}a_j+{\bf u}_j^\mathrm{T}a^\dagger_j+{\bf v}_j^{\mathrm{T}}(s_j-a^\dagger_j a_j))\right) \,,
	\label{hamilton}
\end{align} 
where $a^\dagger_i$/$a_i$ are bosonic operators that decrease/increase the spin quantum number
and $s_i$ is the modulus of the classical spin vector at atomic position $i$.
The vectors ${\bf{u}}_i$ and ${\bf{v}}_i$ are defined in terms of the rotation matrix ${\mathcal R}$
by using the Rodrigues´ formula \cite{rodrigues}
\begin{align}
	u^\beta_i =&{\mathcal R}_i^{\beta 1}+i{\mathcal R}_i^{\beta 2} \,,\nonumber\\
	v^\beta_i =&{\mathcal R}_i^{\beta 3} \,,
\end{align}
where $\beta$ runs over \{$x$,$y$,$z$\}.
The exchange tensor $\mathcal{J}_{ij}$ reads
\begin{align}
\mathcal{J}_{ij}&=J_{ij}\mathcal{I}+\mathcal{J}^S_{ij}+\mathcal{J}^A_{ij}\nonumber\\&=
\begin{pmatrix}
  J_{ij}+\Gamma_{ij}^{xx} & D_{ij}^z  & -D_{ij}^y  \\
  -D_{ij}^z &   J_{ij}+\Gamma_{ij}^{yy} & D_{ij}^x \\
   D_{ij}^y& - D_{ij}^x &   J_{ij}-\Gamma_{ij}^{xx}-\Gamma_{ij}^{yy} 
 \end{pmatrix} \,.
\end{align}
After Fourier transform, the Hamiltonian in Eq.~(\ref{hamilton}) can be recast in the following
form\,:
\begin{align}
	&\mathcal{H}=\nonumber\\
	&\sum_{k\in \mathrm{BZ}}\left(a^\dagger_i({\bf k}) a_i(-{\bf k})\right)
	\begin{pmatrix}
  A({\bf k})-C & B({\bf k}) \\
 B^\dagger({\bf k}) & \bar{A}(-{\bf k})-C    
 \end{pmatrix}
	\begin{pmatrix}
 a_i({\bf k})  \\
    a^\dagger_i({-\bf k}) \nonumber
 \end{pmatrix} \,,
\end{align}
where $A$, $B$, and $C$ are defined as\,:
\begin{align}
	A({\bf k})_{ij}&=\frac{\sqrt{s_is_j}}{2}{\bf u}_i^\mathrm{T}\mathcal{J}_{ij}(-{\bf k})\bar{{\bf u}}_j \,, \label{reciprocal1}\\
	B({\bf k})_{ij}&=\frac{\sqrt{s_is_j}}{2}{\bf u}_i^\mathrm{T}\mathcal{J}_{ij}(-{\bf k}){\bf u}_j \,,\\
	C({\bf k})_{ij}&=\delta_{ij}\sum_l s_l {\bf v}_i^\mathrm{T}\mathcal{J}_{il}(0){\bf v}_l \,.
	\label{reciprocal2}
\end{align}
This is diagonalized by using a Bogoliubov transformation \cite{colpa}.
The calculated eigenvalues are the eigenfrequencies of the spin waves and are plotted in
Figs.~\ref{fig3}, \ref{fig4}.(a)-(b), and S1.

\end{document}